\documentclass[aps,twocolumn,pra,preprintnumbers,showpacs,tightenlines]{revtex4}
\usepackage{amssymb}
\usepackage{amsmath}
\usepackage{graphicx}
\usepackage{epsfig}
\usepackage{subfigure}
\usepackage{amsfonts}
\usepackage{CJK}

\begin{document}

\title{Single-step implementation of a multiple-target-qubit controlled phase gate without need of classical pulses}

\author{Chui-Ping Yang$^{1}$, Qi-Ping Su$^{1}$, Feng-Yang Zhang$^{2}$, and Shi-Biao Zheng$^{3}$}
\email{sbzheng11@163.com}
\address{$^1$Department of Physics, Hangzhou Normal University,
Hangzhou, Zhejiang 310036, China}

\address{$^2$School of Physics and Materials Engineering, Dalian Nationalities University, Dalian 116600, China}

\address{$^3$Department of Physics, Fuzhou University, Fuzhou 350108, China}
\date{\today}

\begin{abstract}
We propose a simple method for realizing a multiqubit phase gate of
one qubit simultaneously controlling $n$ target qubits, by using three-level
quantum systems (i.e., qutrits) coupled to a cavity or resonator. The gate can be implemented using one operational step and
without need of classical pulses, and no photon is populated during the operation. Thus,
the gate operation is greatly simplified and decoherence from the cavity decay is
much reduced, when compared with the previous proposals. In addition, the operation time
is independent of the number of qubits and no adjustment of the qutrit level spacings or the cavity frequency is needed
during the operation.
\end{abstract}
\date{\today}
\maketitle

Multiple qubit gates have many applications in quantum information
processing (QIP). A multiqubit gate can be decomposed into two-qubit and
one-qubit gates and thus can in principle be constructed using these basic
gates. However, the number of basic gates increases dramatically as the
number of qubits increases. Thus, it becomes difficult to build a multiqubit
gate by using the conventional gate-decomposition protocol. Over the past
years, many efficient schemes have been proposed for the direct
implementation of a multiqubit controlled-phase or controlled-NOT gate with
multiple-control qubits acting on one target qubit (e.g., [1-5]). This type
of multiqubit gate plays significant roles in QIP, such as quantum
algorithms and error corrections.

We here focus on another type of multiqubit gates, that is, a multiqubit
phase gate with one control qubit simultaneously controlling multiple target
qubits. This multiqubit phase gate is described by
\begin{eqnarray}
\left| 0_1\right\rangle \left| i_2\right\rangle \left| i_3\right\rangle
...\left| i_n\right\rangle &\rightarrow &\left| 0_1\right\rangle \left|
i_1\right\rangle \left| i_2\right\rangle ...\left| i_n\right\rangle ,
\nonumber \\
\left| 1_1\right\rangle \left| i_1\right\rangle \left| i_2\right\rangle
...\left| i_n\right\rangle &\rightarrow &\left| 1_1\right\rangle \left(
-1\right) ^{i_1}\left( -1\right) ^{i_2}...\left( -1\right) ^{i_n}\left|
i_1\right\rangle  \nonumber \\
&&\;\;\;\;\,\,\left| i_2\right\rangle ...\left| i_n\right\rangle ,
\end{eqnarray}
where the subscript $1$ represents the control qubit while subscripts $%
2,3,...,$ and $n$ represent the $n-1$ target qubits, and $i_2,i_3,...,i_n\in
\{0,1\}.$ The transformation (1) implies that when the control qubit is in
the state $\left| 0\right\rangle ,$ nothing happens to the states of each
target qubit; however, when the control qubit is in the state $\left|
1\right\rangle ,$ a phase flip (from the $+$ sign to the $-$ sign) happens
to the state $\left| 1\right\rangle $ of each target qubit. This multiqubit
gate is useful in QIP such as entanglement preparation, error correction,
and quantum algorithms.

Several methods have been proposed for the direct implementation of this
multiqubit phase gate, by employing two-level or four-level quantum systems
coupled to a single cavity or resonator [6-8]. However, these methods
require several steps of operation and application of classical pulses so
that the gate operation is complex. Moreover, in these schemes cavity
photons are populated during the operation and thus decoherence caused by
the cavity decay may pose a problem. In addition, for the methods proposed
in [7,8], a higher-energy fourth level was employed, which is experimantally
challenging.

In the following, we present a new approach for implementing this multiqubit
phase gate using three-level quantum systems (i.e., qutrits) coupled to a
single cavity or resonator. Compared with the previous proposals, the
proposal has these features: (i) only a single step of operation is needed
and no classical pulse is used, thus the operation is greatly simplified;
(ii) no photon is populated, thus decoherence caused by the cavity photon
decay is much suppressed; and (iii)\emph{\ }it is unnecessary to employ a
fourth level. In addition, the proposal has the following additional
advantages: the operation time is independent of the number of qubits and no
adjustment of the qutrit level spacings or the cavity frequency is required
during the operation.

Consider $n$ qutrits labeled by $1,2,...,$ and $n$. Each qutrit has three
levels $\left| g\right\rangle $, $\left| e\right\rangle $, and $\left|
f\right\rangle $ (Fig. 1). Assume that qutrits $2,3,...,$ and $n$ are
identical, whose levels spacings are different from those of qutrit $1.$ The
cavity mode is coupled to the $\left| e\right\rangle \leftrightarrow \left|
f\right\rangle $ transition of each qutrit, but decoupled (highly detuned)
from the transition between any other two levels (Fig.~1). These
requirements can in principle be met by choosing or designing the qutrits
(e.g., the level spacings of artificial atoms, such as superconducting
quantum devices, can be readily adjusted by varying the device parameters
appropriately). The interaction Hamiltonian in the interaction picture and
under the rotating-wave approximation is given by
\begin{eqnarray}
H_{\mathrm{I}} &=&\hbar \sum_{k=2}^n\mu \left( e^{i\Delta t}a\sigma
_k^{+}+e^{-i\Delta t}a^{+}\sigma _k\right)  \nonumber \\
&&+\hbar \mu _1\left( e^{i\Delta _1t}a\sigma _1^{+}+e^{-i\Delta
_1t}a^{+}\sigma _1\right) ,
\end{eqnarray}
where the subscript $k$ represents the $k$th qutrit, $a^{+}$ ($a$) is the
photon creation (annihilation) operator of the cavity mode with frequency $%
\omega _c$, $\mu $ is the coupling constant between the cavity mode and the $%
\left| e\right\rangle \leftrightarrow \left| f\right\rangle $ transition of
qutrits $\left( 2,3,...,n\right) $, while $\mu _1$ is the coupling constant
between the cavity mode and the $\left| e\right\rangle \leftrightarrow
\left| f\right\rangle $ transition of qutrit $1.$ In addition, $\sigma
_1^{+}=\left| f_1\right\rangle \left\langle e_1\right| $, $\sigma
_k^{+}=\left| f_k\right\rangle \left\langle e_k\right| ,$ $\Delta =\omega
_{fe}-\omega _c$ and $\Delta =\omega _{fe,1}-\omega _c,$ with the $\left|
e\right\rangle \leftrightarrow \left| f\right\rangle $ transition frequency $%
\omega _{fe}$ of qutrits $\left( 2,3,...,n\right) $ and the $\left|
e\right\rangle \leftrightarrow \left| f\right\rangle $ transition frequency $%
\omega _{fe,1}$ of qutrit $1.$

\begin{figure}[tbp]
\begin{center}
\includegraphics[bb=106 311 510 517, width=7.5 cm, clip]{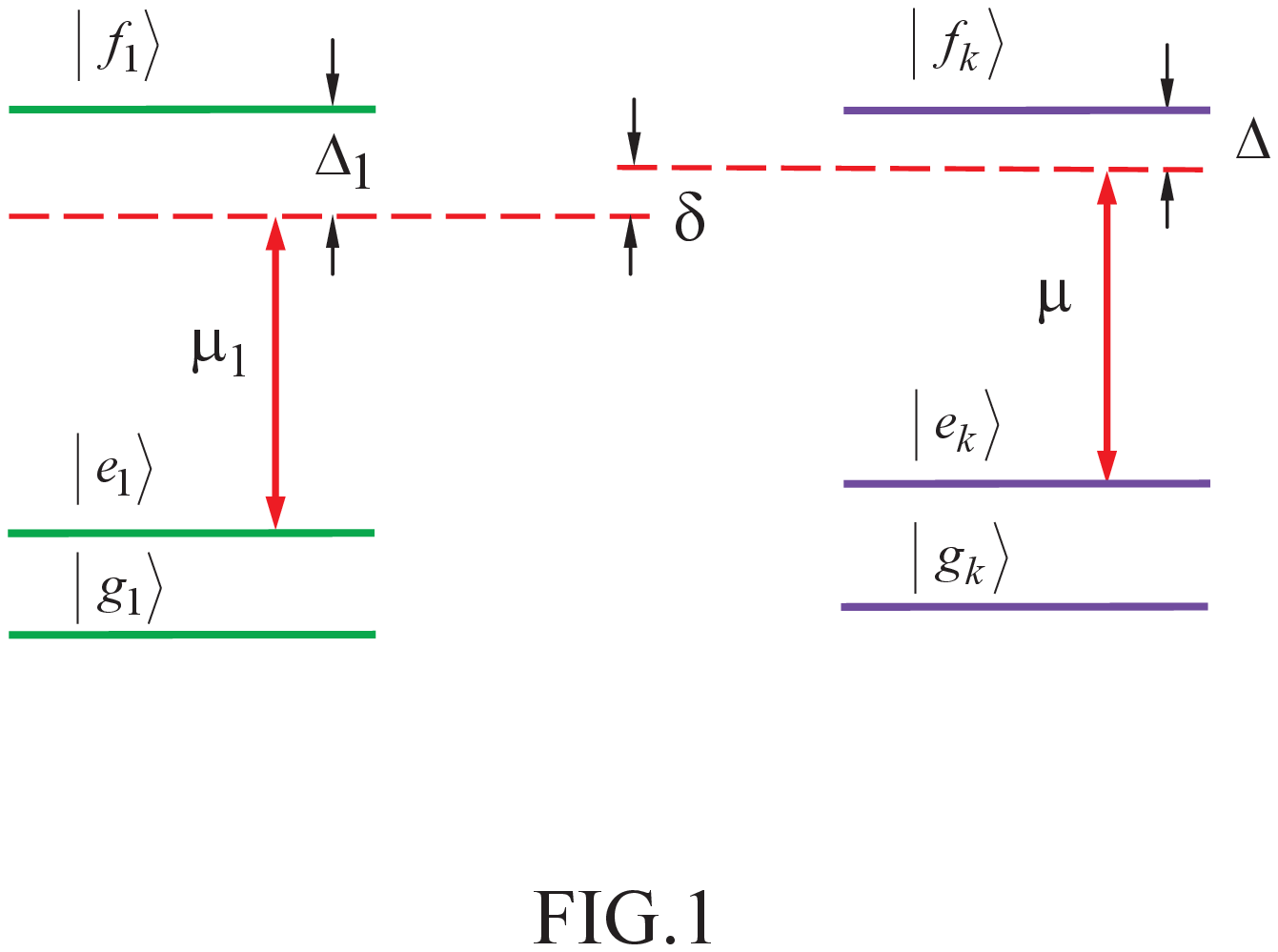} %
\vspace*{-0.08in}
\end{center}
\caption{(Color online) Qutrit-cavity dispersive interaction. The qutrits
could have a $\Lambda $-type, ladder-type or $\Delta$-type three-level
structure. For the $\Lambda $-type, the transition between the two lowest
levels is forbidden or weak. For the ladder-type, the $\left| g\right\rangle
\leftrightarrow \left| f\right\rangle $ transition is forbidden or weak. For
the $\Delta$-type, there exists a transition between any two levels. The
left is for qutrit 1, while the right is for qutrit $k$ ($k=2,3,...,n$).
Note that the level spacing between the two lowest levels can be greater
than that between the two upper levels (not drawn).}
\label{fig:1}
\end{figure}

For $\Delta \gg \mu $ and $\Delta _1\gg \mu _1,$ there is no energy exchange
between the qutrits and the cavity mode. Then the system dynamics described
by the Hamiltonian of Eq. (2) is approximately equivalent to that determined
by the following Hamiltonian [9,10]\emph{\ }
\begin{eqnarray}
H_{\mathrm{I}}^{^{\prime }} &=&-\hbar \sum_{k=2}^n\left[ \frac{\mu ^2}\Delta
\left( a^{+}a\left| e_k\right\rangle \left\langle e_k\right| -aa^{+}\left|
f_k\right\rangle \left\langle f_k\right| \right) \right]   \nonumber \\
&&-\hbar \frac{\mu _1^2}{\Delta _1}\left( a^{+}a\left| e_1\right\rangle
\left\langle e_1\right| -aa^{+}\left| f_1\right\rangle \left\langle
f_1\right| \right)   \nonumber \\
&&\ +\hbar \frac{\mu ^2}\Delta \sum_{k\neq k^{\prime }=2}^n\left( \sigma
_k^{+}\sigma _{k^{\prime }}^{-}+\sigma _k^{-}\sigma _{k^{\prime
}}^{+}\right)   \nonumber \\
&&+\hbar \lambda \sum_{k=2}^n\left( e^{i\delta t}\sigma _1^{+}\sigma
_k^{-}+e^{-i\delta t}\sigma _1^{-}\sigma _k^{+}\right) ,
\end{eqnarray}
where $\lambda =\frac{\mu \mu _1}2\left( \frac 1\Delta +\frac 1{\Delta
_1}\right) $ and $\delta =\Delta _1-\Delta .$ The photon number is conserved
during the interaction. When the cavity mode is initially in the vaccum
state, it will remain in this state. Under this condition the photon number
operator $a^{+}a$ in Eq. (3) can be set to be $0$. Furthermore, when the
condition\emph{\ }$\delta \gg \lambda ,\frac{\mu ^2}\Delta ,\frac{\mu _1^2}{%
\Delta _1}$\emph{\ }is satisfied, qutrit 1 does not exchange energy with the
other qutrits. Under these conditions Hamiltonian (3) can be replaced by the
effective Hamiltonian [11]
\begin{eqnarray}
H_{\mathrm{eff}} &=&\hbar \sum_{k=2}^n\frac{\mu ^2}\Delta \left|
f_k\right\rangle \left\langle f_k\right| +\hbar \frac{\mu _1^2}{\Delta _1}%
\left| f_1\right\rangle \left\langle f_1\right|   \nonumber \\
&&+\hbar \frac{\mu ^2}\Delta \sum_{k\neq k^{\prime }=2}^n\left( \sigma
_k^{+}\sigma _{k^{\prime }}^{-}+\sigma _k^{-}\sigma _{k^{\prime
}}^{+}\right) +\hbar \frac{\lambda ^2}\delta \times   \nonumber \\
&&\sum_{j,k=2}^n\left( \left| f_1\right\rangle \left\langle f_1\right|
\sigma _j^{-}\sigma _k^{+}-\left| e_1\right\rangle \left\langle e_1\right|
\sigma _j^{+}\sigma _k^{-}\right) .
\end{eqnarray}
The last term of Eq. (4) describes the effective coupling of qutrits ($%
2,3,...,n$) arising from the far off-resonant coupling to qutrit 1 described
by the last term of Eq. (3). When the level $\left| f\right\rangle $\ of
each of qutrits ($2,3,...,n$) is not populated, it will remain unpopulated
since the total number of qutrits ($2,3,...,n$) being in the level $\left|
f\right\rangle $ remains unchanged under the effetive Hamiltonian $H_{%
\mathrm{eff}}$.\ In this case, $H_{\mathrm{eff}}$\ reduces to
\begin{eqnarray}
H_{\mathrm{eff}} &=&\hbar \frac{\mu _1^2}{\Delta _1}\left| f_1\right\rangle
\left\langle f_1\right| +\hbar \frac{\lambda ^2}\delta \times   \nonumber \\
&&\sum_{j,k=2}^n\left( \left| f_1\right\rangle \left\langle f_1\right|
\sigma _j^{-}\sigma _k^{+}-\left| e_1\right\rangle \left\langle e_1\right|
\sigma _j^{+}\sigma _k^{-}\right) ,
\end{eqnarray}
which can be further expressed as
\begin{eqnarray}
H_{\mathrm{eff}} &=&\hbar \frac{\mu _1^2}{\Delta _1}\left| f_1\right\rangle
\left\langle f_1\right| +\hbar \frac{\lambda ^2}\delta \left|
f_1\right\rangle \left\langle f_1\right| \sum_{j=2}^n\left| e_j\right\rangle
\left\langle e_j\right|   \nonumber \\
&&+\hbar \frac{\lambda ^2}\delta \left| f_1\right\rangle \left\langle
f_1\right| \sum_{j\neq k;j,k=2}^n\left| e_jf_k\right\rangle \left\langle
f_je_k\right|   \nonumber \\
&&-\hbar \frac{\lambda ^2}\delta \sum_{j,k=2}^n\left( \left|
e_1\right\rangle \left\langle e_1\right| \sigma _j^{+}\sigma _k^{-}\right) .
\end{eqnarray}

We use the asymmetric encoding scheme [5]. Namely, for the gate of Eq.~(1),
the logic state $\left| 0\right\rangle $ of each qubit is represented by the
level $\left| g\right\rangle $, while $\left| 1\right\rangle $\ is
represented by the level $\left| f\right\rangle $\ for qutrit $1$\ but by $%
\left| e\right\rangle $\ for qutrit $j$\ ($j=2,3,...,n$). Since $\left|
e\right\rangle $ is not involved for qutrit $1,$ the last term in Eq.~(6)
can be dropped due to $\left\langle e_1\right| \left. g_1\right\rangle
=\left\langle e_1\right| \left. f_1\right\rangle \equiv 0$. In addition,
because $\left| f\right\rangle $\ is not involved for qutrit $j$\ $\left(
j=2,3,...,n\right) ,$\ the third term in Eq. (6) can be discarded owing to $%
\left\langle f_j\right| \left. g_j\right\rangle =\left\langle f_j\right|
\left. e_j\right\rangle \equiv 0.$\ Hence, the Hamiltonian (6) becomes

\begin{equation}
H_{\mathrm{eff}}=\hbar \frac{\mu _1^2}{\Delta _1}\left| f_1\right\rangle
\left\langle f_1\right| +\hbar \frac{\lambda ^2}\delta \left|
f_1\right\rangle \left\langle f_1\right| \sum_{j=2}^n\left| e_j\right\rangle
\left\langle e_j\right| ,
\end{equation}
for which the time-evolution unitary operator is
\begin{equation}
U=e^{-iH_{\mathrm{eff}}t/\hbar }=U_1\otimes \prod\limits_{j=2}^nU_{1j},
\end{equation}
where $U_1$ is an unitary operator acting on qutrit $1$ while $U_{1j}$ is a
joint unitary operator acting on qutrits $1$ and $j,$ which are given by
\begin{eqnarray}
U_1 &=&\exp \left( -i\frac{\mu _1^2t}{\Delta _1}\left| f\right\rangle
_1\left\langle f\right| \right) , \\
U_{1j} &=&\exp \left( -i\frac{\lambda ^2t}\delta \left| f_1\right\rangle
\left\langle f_1\right| \otimes \left| e_j\right\rangle \left\langle
e_j\right| \right) .
\end{eqnarray}
Note that $U_1\left| g_1\right\rangle =0,$ $U_1\left| f_1\right\rangle =\exp
(-i\mu _1^2t/\Delta _1)\left| f_1\right\rangle ,$ $U_{1j}\left|
g_1\right\rangle \left| l_j\right\rangle =\left| g_1\right\rangle \left|
l_j\right\rangle ,$ and $U_{1j}\left| f_1\right\rangle \left|
l_j\right\rangle =\exp \left( -i\left\langle e_j\right| \left.
l_j\right\rangle \lambda ^2t/\delta \right) \left| f_1\right\rangle \left|
l_j\right\rangle ,$ where $\left| l_j\right\rangle \in \{\left|
g_j\right\rangle ,\left| e_j\right\rangle \}$ ($j=2,3,...,n$). Thus, the
operator $U$ leads to the transformation
\begin{eqnarray}
&&\left| g_1\right\rangle \left| l_2\right\rangle \left| l_3\right\rangle
...\left| l_n\right\rangle \rightarrow \left| g_1\right\rangle \left|
l_1\right\rangle \left| l_2\right\rangle ...\left| l_n\right\rangle ,
\nonumber \\
&&\left| f_1\right\rangle \left| l_2\right\rangle \left| l_3\right\rangle
...\left| l_n\right\rangle \rightarrow e^{-i\mu _1^2t/\Delta _1}\left|
f_1\right\rangle e^{-i\left\langle e_2\right| \left. l_2\right\rangle
\lambda ^2t/\delta }\left| l_2\right\rangle  \nonumber \\
&&\;\;\;\;\;\;\;\;e^{-i\left\langle e_3\right| \left. l_3\right\rangle
\lambda ^2t/\delta }\left| l_3\right\rangle ...e^{-i\left\langle e_n\right|
\left. l_n\right\rangle \lambda ^2t/\delta }\left| l_n\right\rangle .
\end{eqnarray}
For $t=\delta \pi /\lambda ^2=\Delta _12\pi /\mu _1^2,$ i.e., setting $\mu
_1^2/\Delta _1=2\lambda ^2/\delta ,$ the transformation (11) can be further
written as
\begin{eqnarray}
&&\left| g_1\right\rangle \left| l_2\right\rangle \left| l_3\right\rangle
...\left| l_n\right\rangle \rightarrow \left| g_1\right\rangle \left|
l_1\right\rangle \left| l_2\right\rangle ...\left| l_n\right\rangle ,
\nonumber \\
&&\left| f_1\right\rangle \left| l_2\right\rangle \left| l_3\right\rangle
...\left| l_n\right\rangle \rightarrow \left( -1\right) ^{\left\langle
e_2\right| \left. l_2\right\rangle }\left( -1\right) ^{\left\langle
e_3\right| \left. l_3\right\rangle }  \nonumber \\
&&\;\;\;\;\;\;\;\;\,...\left( -1\right) ^{\left\langle e_n\right| \left.
l_n\right\rangle }\left| f_1\right\rangle \left| l_2\right\rangle \left|
l_3\right\rangle ...\left| l_n\right\rangle ,
\end{eqnarray}
which shows that when qutrit $1$ is in the state $\left| g\right\rangle ,$
nothing happens to the states of each of qutrits $\left( 2,3,...,n\right) $;
however, when qutrit $1$ is in the state $\left| f\right\rangle ,$ a phase
flip (from the $+$ sign to the $-$ sign) happens to the state $\left|
e\right\rangle $ of each of qutrits $\left( 2,3,...,n\right) $. Hence, a
multi-target phase gate described by Eq. (1) is realized with $n$ qutrits,
i.e., the control qutrit $1$ and the $\left( n-1\right) $ target qutrits ($%
2,3,...,n$).

To see the above more clearly, let us consider three qutrits for
implementing a three-qubit phase gate. The three-qubit computational basis
corresponds to \{$\left| ggg\right\rangle ,\left| gge\right\rangle ,\left|
geg\right\rangle ,\left| gee\right\rangle ,\left| fgg\right\rangle ,\left|
fge\right\rangle ,\left| feg\right\rangle ,\left| fee\right\rangle $\}.
Based on Eq. (12), one can find that the four states $\left|
fgg\right\rangle ,$ $\left| fge\right\rangle ,$ $\left| feg\right\rangle ,$
and $\left| fee\right\rangle $ of the qutrits become $\left|
fgg\right\rangle ,$ $-\left| fge\right\rangle ,$ $-\left| feg\right\rangle ,$
and $\left( -\right) \left( -\right) \left| fee\right\rangle $,
respectively; while the other four states $\left| ggg\right\rangle ,$ $%
\left| gge\right\rangle ,$ $\left| geg\right\rangle ,$ and $\left|
gee\right\rangle $ remain unchanged.

We now give a general discussion on the fidelity of the operation. After
taking the dissipation and dephasing into account, the dynamics of the lossy
system is determined by the master equation

\begin{eqnarray}
\frac{d\rho }{dt} &=&-i\left[ H_I,\rho \right] +\kappa \mathcal{L}\left[
a\right]  \nonumber \\
&&+\sum_{j=1}^n\left\{ \gamma _j\mathcal{L}\left[ \sigma _j\right] +\gamma
_{j,fg}\mathcal{L}\left[ \sigma _{j,fg}\right] +\gamma _{j,eg}\mathcal{L}%
\left[ \sigma _{j,eg}\right] \right\}  \nonumber \\
&&+\sum_{j=1}^n\gamma _{j\varphi ,f}\left( \sigma _{j,ff}\rho \sigma
_{j,ff}-\sigma _{j,ff}\rho /2-\rho \sigma _{j,ff}/2\right)  \nonumber \\
&&+\sum_{j=1}^n\gamma _{j\varphi ,e}\left( \sigma _{j,ee}\rho \sigma
_{j,ee}-\sigma _{j,ee}\rho /2-\rho \sigma _{j,ee}/2\right) ,  \nonumber \\
&&
\end{eqnarray}
where $H_I$ is the Hamiltonian in Eq. (2), $\sigma _{j,fg}=\left|
g\right\rangle _j\left\langle f\right| ,$ $\sigma _{j,eg}=\left|
g\right\rangle _j\left\langle e\right| ,$ $\sigma _{j,ff}=\left|
f\right\rangle _j\left\langle f\right| ,$ $\sigma _{j,ee}=\left|
e\right\rangle _j\left\langle e\right| ,$ and $\mathcal{L}\left[ \Lambda
\right] =\Lambda \rho \Lambda ^{+}-\Lambda ^{+}\Lambda \rho /2-\rho \Lambda
^{+}\Lambda /2,$ with $\Lambda =a,$ $\sigma _j,$ $\sigma _{j,fg},$ and $%
\sigma _{j,eg}$.\emph{\ }$\kappa $\emph{\ }is the photon decay rate of the
cavity$,$ $\gamma _{j,eg}$ is the relaxation rate of the level $\left|
e\right\rangle $ of qutrit $j$ for the decay path $\left| e\right\rangle
\rightarrow \left| g\right\rangle $, $\gamma _{j,fg}$ ($\gamma _j$) is the
relaxation rate of the level $\left| f\right\rangle $ of qutrit $j$ for the
decay path $\left| f\right\rangle \rightarrow \left| g\right\rangle $ ($%
\left| e\right\rangle $), and $\gamma _{j\varphi ,f}$ ($\gamma _{j\varphi
,e} $) is the dephasing rate of the level $\left| f\right\rangle $ ($\left|
e\right\rangle $) of qutrit $j$ ($j=1,2,...,n$)$.$

\begin{figure}[tbp]
\begin{center}
\includegraphics[bb=0 0 1000 600, width=8.5 cm, clip]{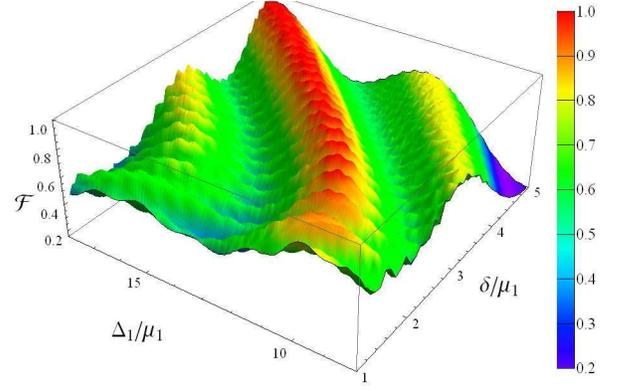} %
\vspace*{-0.08in}
\end{center}
\caption{(Color online) Fidelity versus $\Delta _1/\mu _1$\ and $\delta /\mu
_1$, without considering the system dissipation and dephasing.}
\end{figure}

The fidelity of the operation is given by $\mathcal{F}=\left\langle \psi
_{id}\right| \widetilde{\rho }\left| \psi _{id}\right\rangle ,$ where $%
\left| \psi _{id}\right\rangle $ is the output state for an ideal system
(i.e., without dissipation and dephasing) after the entire operation, while $%
\widetilde{\rho }$ is the final density operator of the whole system when
the operation is performed in a realistic physical system.

For the sake of definitiveness, let us consider the experimental feasibility
of realizing a three-qubit phase gate. Assume that qutrit $1$ is in the
state $\left( \left| g\right\rangle +\left| f\right\rangle \right) /\sqrt{2}%
, $ qutrits $2$ and $3$ are in $\left( \left| g\right\rangle +\left|
e\right\rangle \right) /\sqrt{2},$ and the cavity mode is in the vacuum
state before the gate operation.

\begin{figure}[tbp]
\begin{center}
\includegraphics[bb=0 0 700 400, width=8.0 cm, clip]{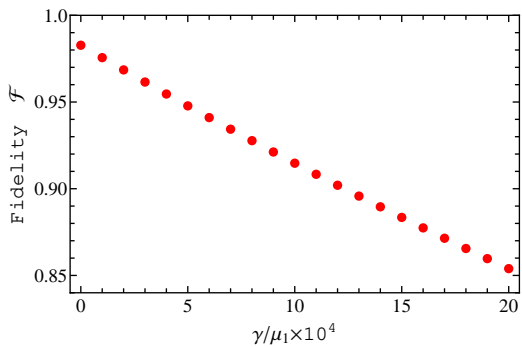} %
\vspace*{-0.08in}
\end{center}
\caption{(Color online) Fidelity as a function of $\gamma/\mu _1$, after
taking the system dissipation and dephasing into account.}
\label{fig:3}
\end{figure}

Fig. 2 is plotted for the fidelity versus $\Delta _1/\mu _1$\ and $\delta
/\mu _1$, without considering the dissipation and dephasing of the whole
system. The middle red-color convex surface in Fig. 2 shows that a high
fidelity $\geq 99\%$\ can be achieved for a wide range of $\Delta _1/\mu _1$%
\ and $\delta /\mu _1$. Without loss of generality, choose $\Delta
_1=10.7\mu _1,$\ $\Delta =8.4\mu _1$, and $\mu =3.08\mu _1$. With these
parameters and the dissipation and dephasing considered, we solve the master
equation numerically. As an example, consider qutrits with a ladder-type
level structure (available in natural atoms, quantum dots, superconducting
phase qutrits, and transmon qutrits). We set $\kappa =0.01\mu _1$, $\gamma
_{j\varphi ,e}=\gamma _{j\varphi ,f}=\gamma _j=\gamma _{j,eg}=\gamma $, and $%
\gamma _{j,fg}=0.01\gamma $, and plot the fidelity as a function of $\gamma
/\mu _1$\ in Fig. 3. The result shows that when $\gamma /\mu _1\leq 2\times
10^{-4},$\ a fidelity higher than $96.8\%$\ can be obtained. The fidelity
can be further increased by optimizing the system parameters. For $\gamma
/\mu _1=2\times 10^{-4}$\ and $\mu _1=2\pi \times 85$\ MHz$,$\ the qutrit
decoherence time is $9.36$\ $\mu $s and $\mu =2\pi \times 261.8$\ MHz, which
are readily available for superconducting transom qutrits. Decoherence time
can be made to be on the order of $20-60$\ $\mu $s for state-of-the-art
superconducting transom devices [12-14], and a coupling constant $\sim 2\pi
\times 360$\ MHz has been reported for a superconducting transmon device
coupled to a resonator [15]. For superconducting qutrits, the typical
transition frequency between two neighbor levels is between $5$\ and $10$\
GHz. As an example, we take $\omega _{fe,1}/2\pi =6.0$\ GHz, $\omega _c/2\pi
=5.09$\ GHz, $\mu _1=2\pi \times 85$\ MHz, and $\kappa =2\pi \times 0.85$\
MHz. The corresponding quality factor of the resonator is $Q$\ $=5.97\times
10^3$\ (a value much lower than $Q$\ $\sim 10^5$\ required by [6,7]). Note
that superconducting resonators with a loaded quality factor $Q\sim 10^6$\
have been experimentally demonstrated [16,17].

In conclusion, we have presented an approach for implementing the multiqubit
phase gate. As shown above, the operation is greatly simplified and
decoherence caused by the cavity photon decay is much reduced when compared
with the previous proposals. This work is quite general, and can be applied
to a wide range of physical implementation with natural atoms or artificial
atoms (e.g., quantum dots, NV centers, or various superconducting qutrits
such as flux, phase, charge, and transmon qutrits) coupled to a cavity or
resonator.

This work was supported by the Major State Basic Research Development
Program of China under Grant No. 2012CB921601, the National Natural Science
Foundation of China under Grant Nos. [11074062, 11374083, 11147186,
11175033, 11374054] and the Zhejiang Natural Science Foundation under Grant
No. LZ13A040002. C.P.Y. acknowledges the funding support from Hangzhou
Normal University under Grant Nos. HSQK0081 and PD13002004 and Hangzhou City
for the Hangzhou-City Quantum Information and Quantum Optics Innovation
Research Team.

\end{document}